# Visualizing the melting of the charge density wave in UTe$_2$ by generation of pairs of topological defects with opposite winding

Anuva Aishwarya[1], Julian May-Mann[1,2], Avior Almoalem[1], Sheng Ran[3,4,5], Shanta R. Saha[3], Johnpierre Paglione[3,6], Nicholas P. Butch[3,4], Eduardo Fradkin[1,2], Vidya Madhavan[1,6]*

[1]Department of Physics and Materials Research Laboratory, University of Illinois at Urbana-Champaign, Urbana, IL, USA.

[2]Institute for Condensed Matter Theory, University of Illinois, 1101 West Springfield Avenue, Urbana, Illinois 61801, USA

[3]Maryland Quantum Materials Center, Department of Physics, University of Maryland, College Park, MD, USA

[4]NIST Center for Neutron Research, National Institute of Standards and Technology, Gaithersburg, MD, USA.

[5]Department of Physics, Washington University in St. Louis, St. Louis, MO 63130, USA

[6]Canadian Institute for Advanced Research, Toronto, ON M5G 1Z8, Canada

*Correspondence to: vm1@illinois.edu

**Abstract:**

**Topological defects are singularities in an ordered phase that can have a profound effect on phase transitions and serve as a window into the order parameter. In this work we use scanning tunneling microscopy to visualize the role of topological defects in the novel magnetic field induced disappearance of an intertwined charge density wave (CDW) in the heavy fermion superconductor, UTe$_2$. By simultaneously imaging the amplitude and phase of the CDW order, we reveal pairs of topological defects with positive and negative phase winding. The pairs are directly correlated with a zero CDW amplitude and increase in number with increasing magnetic field. These observations can be captured by a Ginzburg Landau model of a uniform superconductor coexisting with a pair density wave. A magnetic field generates vortices of the superconducting and pair density wave order which can create topological defects in the CDW and induce the experimentally observed melting of the CDW at the upper critical field. Our work reveals the important role of magnetic field generated topological defects in the melting the CDW order parameter in UTe$_2$ and provides support for the existence of a parent pair density wave order on the surface of UTe$_2$.**

Ordered phases that break continuous symmetries such as superconductors, magnets, liquid crystals and charge density wave systems, can host special topological defects (1-3) such as vortices (4,5), skyrmions (6-8), disclinations (1-3) and dislocations (9-11). In general, while a defect is a simple phase slip in the order parameter (OP), a topological defect is one which cannot be repaired by a continuous OP change. Such defects are associated with a phase winding of the OP in integer multiples of $2\pi$ as shown in Figure 1d, with the OP going to zero at the defect center. The detailed behavior of topological defects can be used as a window into the nature of the ordered phase. A famous example is the case of superfluid $^3$He where the sub-phases in the superfluid state can be classified by the nature of the superconducting vortices (12,13). Additionally, in contrast to the usual melting of an OP via a continuous suppression of the amplitude, the generation of topological defects can play a critical role in phase transitions. The latter occurs in the famous BKT transition (14-18) in two-dimensions, where the long-range order of the XY model is destroyed by the thermal generation and proliferation of pairs of topological defects with opposite winding. Similarly, applying a magnetic field to a type II superconductor leads to the creation of vortices, the number of which increases as the magnetic field is increased, until the superconductivity is destroyed at the upper critical field, $H_{c2}$ (4,5). Phase transitions through the generation of topological defects appears not only in condensed matter physics, but also in other areas of physics like studies of the early universe (19).

UTe$_2$ is a heavy fermion superconductor that exhibits an extremely rich phase diagram with pressure, temperature, and magnetic-field (20,21) including a low-temperature, magnetic-field sensitive charge order on its surface which is strongly intertwined with superconductivity (22). In this work, we employ high resolution scanning tunneling microscopy and spectroscopic imaging to visualize the generation of pairs of topological defects with opposite phase winding that are responsible for the magnetic field induced melting of the unconventional CDW in UTe$_2$ (22). UTe$_2$ crystallizes into a body-centered orthorhombic structure with the space group *Immm* as shown in Figure 1a. The crystal has two types of Te atoms based on the relative U-Te bond lengths as indicated by the Figure 1a. UTe$_2$ is a Kondo metal below about 30 K and superconducts below 1.6 K (refer to Supplementary Figure 1a for transport characterization). UTe$_2$ single crystals used in this study were cleaved at ~ 90 K in ultrahigh vacuum and immediately inserted into the STM head. The (011) plane of the crystal is the easy cleave plane and cleaving along this plane exposes chains of Te1 and Te2 atoms that align along the crystallographic a-direction as indicated by the schematic in Figure 1b. Figure 1c shows an atomically resolved STM topography with the schematic of the atomic sites overlayed. Shown in Figure1e, is a large area atomically resolved STM topography obtained at 300 mK. Figure 1f is the Fourier transform (FT) corresponding to the topography shown in Figure 1e, where the Bragg peaks corresponding to the Te-lattice ($q^{Te}_{1,2}$) and the additional modulations corresponding to incommensurate CDWs ($q^{CDW}_{1,2,3}$), are labelled (22). The CDW order parameter has an amplitude and phase, both of which can vary as a function of position in real space. The three incommensurate CDW orders in UTe$_2$ can be expressed mathematically as $\rho_{q_i}(r) = \rho^o_{q_i}(r)cos(q_i r + \phi_i(r)), i = 1,2,3$, where $\rho^o_{q_i}(r)$ is the amplitude of the $i$th CDW, and $cos(q_i r + \phi_i(r))$ captures the periodic modulation of the CDW. We refer to $\phi_i(r)$ as the relative phase of the CDW to distinguish it from $q_i r$. The relative phase of the CDW can vary in real space (due to dislocations for example). Our goal in this work is to use the information contained in the amplitude, $\rho^o_{q_i}(r)$ and relative phase, $\phi_i(r)$ to determine the mechanism of the magnetic field induced melting of the CDWs in UTe$_2$.

The technique to determine the spatial variations in the amplitude and relative phase of each of the CDWs is the following. We first perform an inverse Fourier filtering of the CDW peaks in the FT to isolate the signal associated with the CDW modulation. To visualize the amplitude and the modulating component for the CDW in real space, we take the modulus and the cosine of the argument of this signal and plot those as a function of position ($r$). More details of this process can be found in the supplement. To illustrate the types of data sets and the information contained in them, we use this technique to obtain the three quantities $\rho^o_{q_1}(r)$, $cos(q_1 r + \phi_1(r))$, and $\phi_1(r)$ for q$_1^{CDW}$ (labelled in the FT Figure 1c). The resulting images are shown in Figures 1d-f. A similar analysis performed for q$_2^{CDW}$ is shown in Supplementary Figure 3. At zero magnetic field, we observe very few dislocations of the CDW consistent with long range CDW order. However, some regions have dislocations as shown in Figure 1e (and Supplementary Figure 3 and 4). The dislocations are topological defects in the CDW order and can be seen as a phase winding in in the relative phase ($\phi_1(r)$) map shown in Figure 1f (and Supplementary Figure 3 and 4).

Prior work has shown that the CDW OP is suppressed with magnetic field and disappears close to H$_{c2}$ (22), and our goal is to determine the role of topological defects in this phase transition. Figure 2a-d show a series of FTs of topographies with increasing field. As shown previously, the CDW peaks in the FTs are suppressed and eventually disappear close to H$_{c2}$. We first look at the evolution of the modulating component of q$_2^{CDW}$. As the magnetic field increases, we find an increasing number of topological defects (indicated by circles in Figure 2e-h). This holds true for q$_1^{CDW}$ and q$_3^{CDW}$ as well, as shown in Supplementary Figure 5 and 6 e-h. The primary conclusion from these maps is that an increasing magnetic field leads to an increase in the number of dislocations in all components of the CDW. However, the positions of the dislocations in $\rho^o_{q_1}(r)$, $\rho^o_{q_2}(r)$, and $\rho^o_{q_3}(r)$ are not necessarily the same. We will come back to this point later in the paper.

We now turn our attention to the amplitude maps which show a steady decrease in the overall amplitude for all components of the CDW as function of field (Figure 2i-l, all plotted on the same intensity scale, also Supplementary Figure 5 a-d and 6a-d for $q_2^{CDW}$ and $q_3^{CDW}$ respectively). This is clearly reflected in the histograms of the amplitude values (Figure 2m-p) which show that the mean value of the distribution (indicated by the red dashed line) goes down with increasing field.

The natural next question is whether the amplitude suppression with magnetic field is caused by the increasing number of topological defects, or if they are independent. In fact, the CDW amplitude necessarily goes to zero at the center of the topological defect. Could the generation of these topological defects be the mechanism by which the CDW is destroyed? To answer this question, we create a `zero-amplitude' map i.e., a map of areas where the amplitude is within a small range of zero (Figure 3a-b) and compare this with a map of the locations of the topological defects in same field of view (Figure 3c-d). The `zero-amplitude' map, was created with a binary mask for the zero-amplitude region using a 5% thresholding. Comparing Figure 3c and 3d, we find that the zero-amplitude areas and the sites of the topological defects have a direct one-to-one correspondence. This is borne out by the strong positive cross-correlation between the two maps (value of 0.6, where +1/-1 is very strong positive/negative correlation in Figure 3e). A similar analysis performed for $q_1^{CDW}$ shows a cross-correlation value of 0.7 (Supplementary Figure 7). Our data indicate that the amplitude $\rho^o_{q_i}$ does not diminish uniformly in real space. Instead, the magnetic field generates topological defects which punch holes in the amplitude field. These data provide compelling evidence for the magnetic field induced melting of the CDW OP in UTe$_2$ through the generation of an increasing number of topological defects.

The description above bears marked resemblance to the vortex state of a type-II superconductor where the superconducting order parameter goes to zero inside vortex cores, while still being finite outside it. Vortices are topological defects where the phase of the order parameter winds by multiples of $2\pi$, and can occur in any ordered phase. To visualize this phase winding around a CDW dislocation, we plot the relative phase as a function of angle for two different dislocations as shown in Figure 4. We find that in a non-zero magnetic field, the CDW hosts both dislocations and anti-dislocations with opposite vorticities (where the relative phase winds by $+2\pi$ and $-2\pi$, respectively Figure 4e and 4f respectively). A polar plot far away from the dislocation shows a constant relative phase as expected (Figure 4g). Dislocation and anti-dislocations are both expected since there is no preferred orientation for the Burger's vectors (provided they are parallel to the CDW wavevector). This is unlike superconducting vortices generated by a magnetic field, which have a preferred winding set by the direction of the magnetic field. Interestingly, an examination of the phase maps with increasing magnetic field shows that the defects occur in dislocation-anti-dislocation pairs (shown by pairs of black and white circles in Figure 4e-g) which prefer to remain close to each other. This makes energetic sense, as net vorticity costs extra energy, and dislocations of opposite vorticity can minimize energy by pairing up.

Previous work (22) has suggested that this unconventional CDW in UTe$_2$ is strongly intertwined (23) with a pair density wave (PDW) and uniform superconductivity. Additionally, Josephson STM measurements have provided evidence for a PDW with same wavevectors as the CDWs on the surface in UTe$_2$ (24), which is also confirmed by our spectroscopic measurements (see Supplementary Figure 8). A PDW state is an exotic phase of matter, a Larkin-Ovchinnikov-like state (25,26) with finite momentum Cooper pairs. To further understand the connection between the three aforementioned orders and the mechanism behind the magnetic field induced dislocations, we consider the Ginzburg-Landau free energy for a system with coexisting uniform triplet superconductivity and triplet PDWs. As previously established, this combination of superconducting orders leads to a daughter CDW order, with the same wavevector as the PDW. Here, we shall also show that half-vortices of the PDW (elementary defects of the PDW where the phase of $\vec{\Delta}_{+q_i}$ winds by $2\pi$ *or* the phase of $\vec{\Delta}_{-q_i}$ winds by $2\pi$, but not both (27)) can also lead to dislocations and anti-dislocations in the CDW (28).

We first show that half-vortices of the PDW can induce dislocations of the CDWs within the Ginzburg Landau theory. Here, we will rewrite the real CDW order parameter, $\rho_{q_i}$ as two complex components with opposite momentum, $\rho_{q_i} = \rho_{+q_i} e^{iq_i \cdot r} + \rho_{-q_i} e^{-iq_i \cdot r}$ with $\rho_{-q_i} = \rho^*_{+q_i}$. Within this theory, there is a trilinear coupling term linking the three OPs ($\rho_{+q_i}$, $\vec{\Delta}_{\pm q_i}$, and $\vec{\Delta}_0$). The Landau theory equations of motion for the CDW order parameter $\rho_{+q_i}$, are

$$\rho_{+q_i} \propto \vec{\Delta}_0 \cdot \vec{\Delta}^*_{-q_i} + \vec{\Delta}^*_0 \cdot \vec{\Delta}_{+q_i}. \quad (1)$$

The equations of motion for $\rho_{-q_i}$ are found by complex conjugation. Dislocations and anti-dislocations are encoded as phase windings of $\rho_{+q_i}$. As we shall now show, a half-vortex of $\vec{\Delta}_{+q_i}$ can lead to a dislocation of $\rho_{+q_i}$ (positive phase winding) and a half-vortex of $\vec{\Delta}_{-q_i}$ can lead to an anti-dislocation of $\rho_{+q_i}$ (negative phase winding). Let us first consider a half-vortex of $\vec{\Delta}_{+q_i}$ where $\vec{\Delta}_{+q_i} \propto e^{i\theta_v}$; $\theta_v$ winds by $2\pi$ around the core of the half-vortex; and $\vec{\Delta}_0$, and $\vec{\Delta}_{-q_i}$ are constant. If $|\vec{\Delta}^*_0 \cdot \vec{\Delta}_{+q_i}| > |\vec{\Delta}_0 \cdot \vec{\Delta}^*_{-q_i}|$ away from the half-vortex core, the half-vortex of $\vec{\Delta}_{+q_i}$ will lead to a dislocation of $\rho_{+q_i}$. However, if $|\vec{\Delta}_0^* \cdot \vec{\Delta}_{+q_i}| < |\vec{\Delta}_0 \cdot \vec{\Delta}_{-q_i}^*|$ there will not be a dislocation. Similar reasoning indicates that a half-vortex of $\vec{\Delta}_{-q_i}$ will lead to an anti-dislocation of $\rho_{+q_i}$, if $|\vec{\Delta}_0 \cdot \vec{\Delta}_{-q_i}^*| > |\vec{\Delta}_0^* \cdot \vec{\Delta}_{q_i}|$ away from the half-vortex core. It is also possible that a vortex of the uniform superconducting order parameter, $\vec{\Delta}_0$, can produce dislocations or anti-dislocations, but this is not expected as discussed in the supplement.

Let us now consider the effect of an external magnetic field. Since $\vec{\Delta}_0$ or $\vec{\Delta}_{\pm q_i}$ are all charge 2e order parameters, a magnetic field will induce vortices of $\vec{\Delta}_0$ and half-vortices of $\vec{\Delta}_{\pm q_i}$. Two PDW half-vortices must also accompany each vortex of the uniform component due to the phase locking that arises from the quartic terms proportional to $\vec{\Delta}_0 \cdot \vec{\Delta}_0 \vec{\Delta}_{+q_i}^* \cdot \vec{\Delta}_{-q_i}^*$ and $\vec{\Delta}_0 \cdot \vec{\Delta}_{+q_i}^* \vec{\Delta}_0 \cdot \vec{\Delta}_{-q_i}^*$ in the Ginzburg-Landau theory. If the cores of vortex and accompanying half-vortices all occupy the same location, they will not induce any dislocations since the phase windings on the right-hand side of Eq. (1) all cancel. However, if the vortex and half-vortex cores all occur at different locations (i.e., if the vortices and half-vortices all repel each other), then there will be local regions where there is only a single half-vortex. Based on our previous arguments, such isolated half-vortices may induce dislocations and anti-dislocations of $\rho_{q_i}$. Since the half-vortices of different PDW components repel each other and they can occur at different locations. This would cause the dislocations of different CDW components to occur at different locations, consistent with our data. The number of these half-vortices will increase as the magnetic field increases, which, in turn, increases the number of dislocations in the CDW. This eventually leads to the full destruction of the CDW.

In summary, we show that the sensitivity of the CDW to magnetic fields in UTe$_2$ is caused by field generated topological defects, i.e., dislocations which suppress the CDW order parameter. While the defects can generally have positive or negative phase winding, we find that they occur in pairs with opposite winding due to energetic considerations. Our analysis also provides important information on the relationship of the CDW to a coexisting PDW state. There are two kinds of PDWs that may occur. In one case, the PDW is the daughter phase arising from a combination of an existing CDW state and superconductivity (29). Here, even if superconductivity is destroyed with field, the CDW will continue to survive. In the second scenario, the PDW is the parent phase with the CDW being the daughter order (23). In this case, the CDW will disappear when either the PDW or uniform SC order are destroyed. Within the PDW scenario, based on disappearance of the CDW along with superconductivity, our data and theoretical analysis suggest that the PDW is the parent phase. Given the surmounting experimental evidence pointing towards an exotic superconducting state in UTe$_2$, this observation potentially makes UTe$_2$ a promising candidate to explore properties of a parent PDW order on its surface.

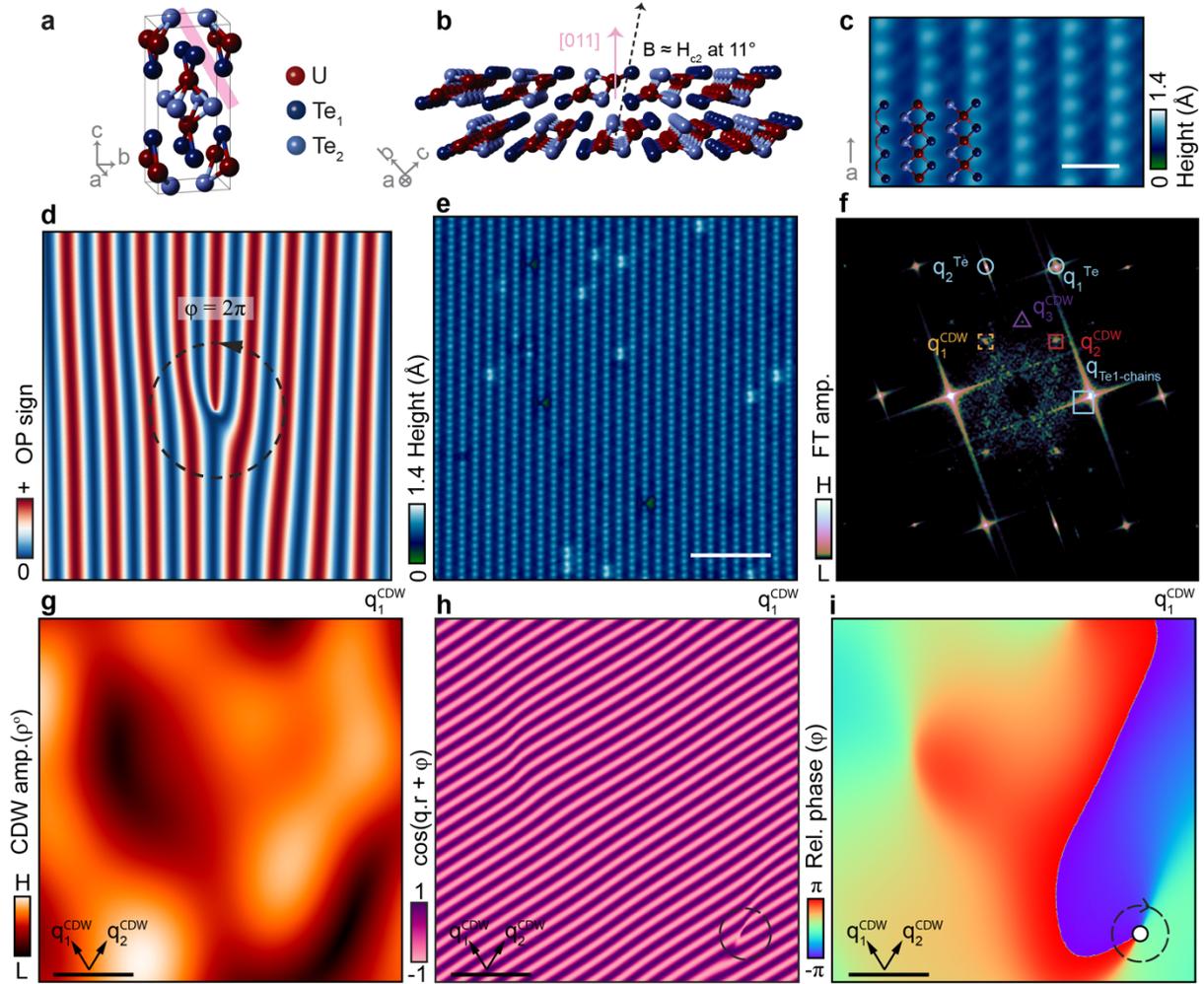

**Figure 1: Topological defects in the CDW in UTe$_2$ and their relationship to the amplitude and phase of the CDW**

**a,** Unit cell of UTe$_2$ with the (011) cleave plane shown by the pink rectangle. **b,** Schematic of the (011) direction and the relative orientation of the applied magnetic field with respect to the (011) direction used in this study. c, Atomically resolved topography of the surface of UTe$_2$ showing the Te1 and Te2 atoms, with an overlay of the atoms of cleave plane as a guide to the eye. Scale bar is 10 Å. **d,** Schematic of a dislocation or a topological defect in an ordered phase where the phase winds by 2π. **e,** High resolution topography on the (011) surface of UTe$_2$ showing the Te1 and Te2 atoms (V = -40 mV, I = 120 pA, T = 300 mK). Scale bar is 50 Å. **f,** Fourier transform of the topography shown in **e**, with the lattice Bragg peaks and the CDW peaks marked. **g-i,** Amplitude, $cos(q_1 r + \phi_1(r))$ and relative phase maps of the order parameter associated with q$_1^{CDW}$. Scale bars are 50 Å. **h** and **i** show the position of an isolated topological defect in the CDW. The topography and various maps shown here have been cropped and aligned from a larger field of view for ease of visualization. The original topography and relative phase maps for both components of the CDW have been provided in Supplementary Figure 2 and 3.

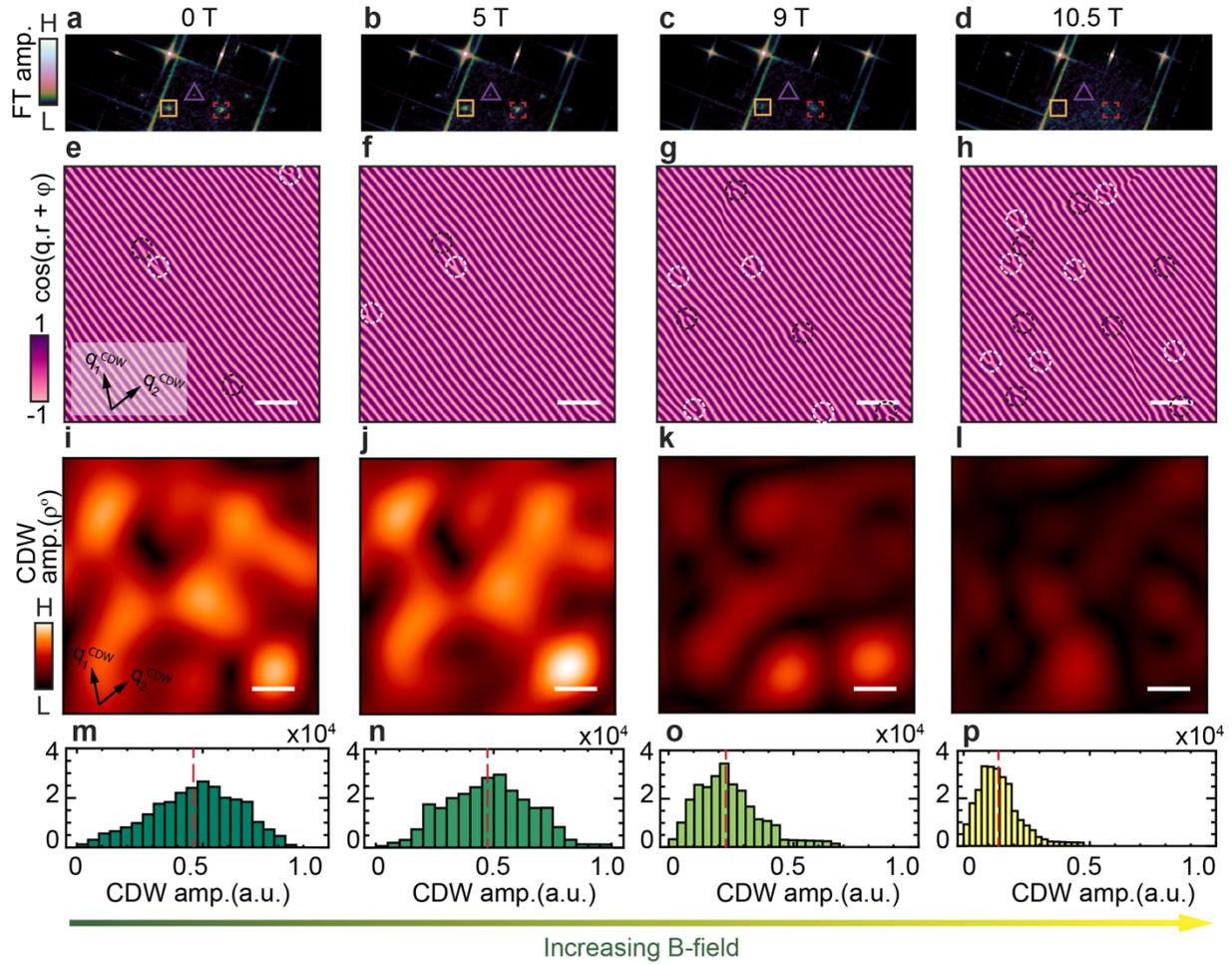

**Figure 2: Generation of topological defects in the phase of the CDW and decay of the amplitude with increasing magnetic field**

**a-d**, Fourier transform of topographies obtained on the same area as a function of increasing magnetic field at T = 300 mK. The CDW peaks marked by squares and triangle. $q_2^{CDW}$ peak which is used to obtain maps of the amplitude and modulating part is shown by the dashed square. **e-h**, $cos(q_2 r + \phi_2(r))$ maps of $q_2^{CDW}$ as a function increasing magnetic field showing an increasing number of dislocations and anti-dislocations which are indicated by black and white dashed circles respectively. The CDW directions have been marked in **e** as a reference. **i-l**, Suppression of amplitude shown by amplitude maps of $q_2^{CDW}$ as a function increasing magnetic field in the same field of view. The CDW directions have been marked in **i** as a reference. **m-p**, Histogram of the amplitude maps of $q_2^{CDW}$ as a function increasing magnetic field where the mean amplitude is indicated by red dashed line. All scale bars are 50 Å.

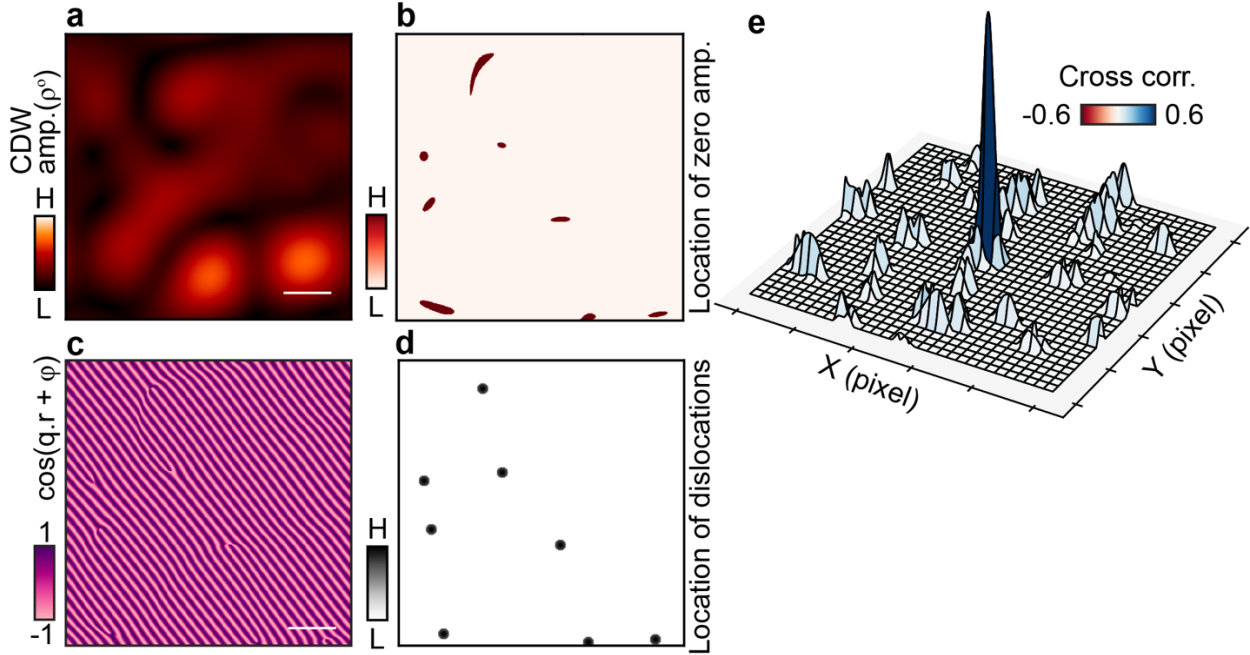

**Figure 3: Strong positive cross-correlation between regions of zero-amplitude and location of the topological defects**

**a**, Amplitude map of $q_2^{CDW}$ obtained at B = 9 T, T = 300 mK. **b**, Zero-amplitude mask generated by a 5% thresholding from the amplitude map which shows regions of zero amplitude. All scale bars are 50 Å. **c**, $cos(q_2 r + \phi_2(r))$ map of $q_2^{CDW}$ on the same field of view as **a**. **d**, Binary mask of the position of dislocations obtained from **c**. Each dislocation is represented by a circular gaussian function. **e**, 2D cross-correlation of the normalized masks shown in **b** and **d** with a value 0.6 at the center, indicative of a positive cross-correlation. A similar cross-correlation value is obtained on analyzing $q_1^{CDW}$ which is shown in the Supplementary Figure 7.

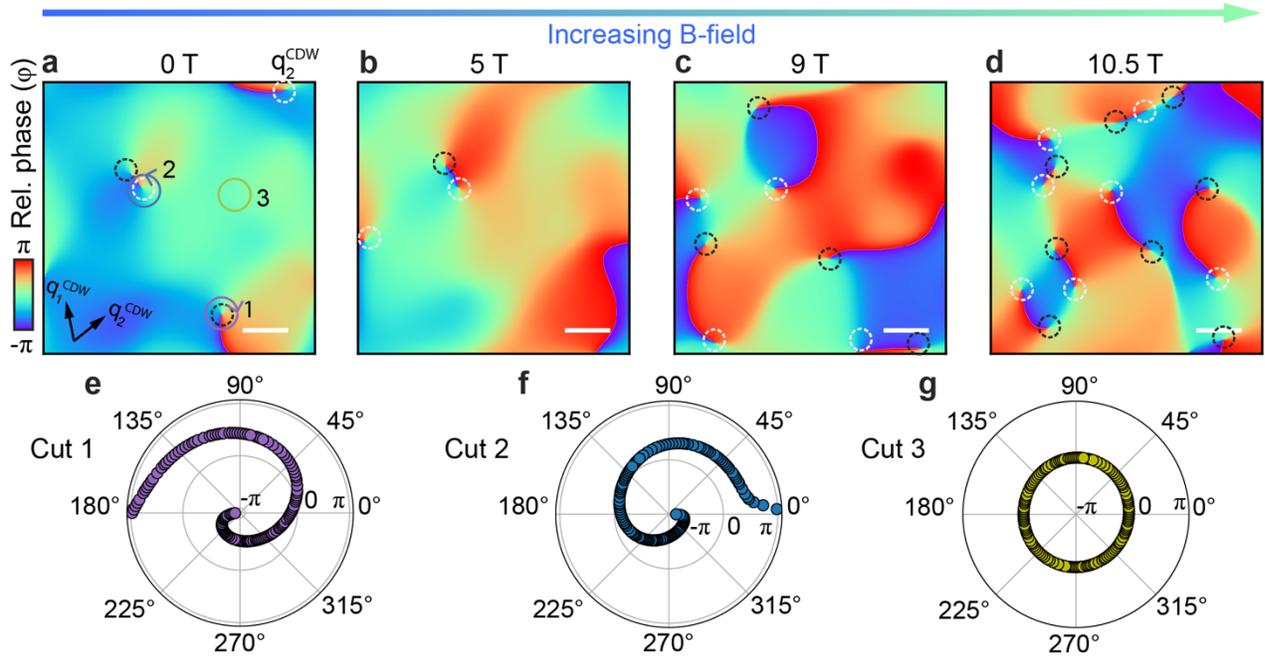

**Figure 4: Increase in pairs of dislocations of the CDW with opposite vorticities as a function of increasing magnetic field.**

**a-d,** Relative phase ($\phi_2(r)$) maps of $q_2^{CDW}$ as a function of magnetic field showing increasing numbers of pairs of dislocations with opposite windings as indicated by the black and white dashed circles respectively. **e-g**, Polar plots around a dislocation, anti-dislocation and a dislocation free area (indicated by 1, 2 and 3 respectively in **a**) showing the phase winding of the CDW OP as a function of angle. The dislocation and anti-dislocation (1 and 2) have opposite winding of the phase, whereas the phase is constant for 3. The phase winds by ±2π for 1 and 2. The plots have been obtained at the same, fixed radius from the defects.

**Acknowledgements**

The authors thank Steve Kivelson, Eun-ah Kim, and Daniel Agterberg for useful discussions. The authors would also like to thank Dr. Ian Hayes who provided the transport characterization of the crystals during the review process. STM work at the University of Illinois, Urbana-Champaign was supported by the U.S. Department of Energy (DOE), Office of Science, Office of Basic Energy Sciences (BES), Materials Sciences and Engineering Division under Award No. DE-SC0022101. V.M. and J.P. acknowledge support from the Gordon and Betty More Foundation's EPiQS Initiative through grants GBMF4860 and GBMF9071 respectively, as well as the Canadian Institute for Advanced Research Quantum Materials Program. Theoretical work was supported in part by the US National Science Foundation through the grant DMR 2225920 at the University of Illinois (E.F.). Research at the University of Maryland was supported by the Department of Energy Award No. DE-SC-0019154 (sample characterization), the National Science Foundation under Grant No. DMR-2105191 (sample preparation), the Maryland Quantum Materials Center, and the National Institute of Standards and Technology. S.R.S. acknowledges support from the National Institute of Standards and Technology Cooperative Agreement 70NANB17H301.

**Author contributions**

A.A. and V.M. conceived the experiments. The single crystals were provided by S.R., S.R.S., J.P. and N.P.B. A.A. and A.A. obtained the STM data. A.A. and V.M. performed the analysis and J.M.M. and E.F. provided the theoretical input on the interpretation of the data. A.A., V.M., J.M.M. and E.F. wrote the paper with input from all authors.

**Competing Interests**

The authors declare no competing interests.

**Supplementary information for**

**Visualizing the melting of the charge density wave in UTe$_2$ by generation of pairs of topological defects with opposite winding**

Anuva Aishwarya[1], Julian May-Mann[1,2], Avior Almoalem[1], Sheng Ran[3,4,5], Shanta R. Saha[3], Johnpierre Paglione[3,6], Nicholas P. Butch[3,4], Eduardo Fradkin[1,2], Vidya Madhavan[1,6]*

[1]Department of Physics and Materials Research Laboratory, University of Illinois at Urbana-Champaign, Urbana, IL, USA.
[2]Institute for Condensed Matter Theory, University of Illinois, 1101 West Springfield Avenue, Urbana, Illinois 61801, USA
[3]Maryland Quantum Materials Center, Department of Physics, University of Maryland, College Park, MD, USA
[4]NIST Center for Neutron Research, National Institute of Standards and Technology, Gaithersburg, MD, USA.
[5]Department of Physics, Washington University in St. Louis, St. Louis, MO 63130, USA
[6]Canadian Institute for Advanced Research, Toronto, ON M5G 1Z8, Canada

1. Methods
2. Supplementary figures 1-8
3. Ginzburg-Landau theory for coexisting uniform triplet superconductivity and triplet pair density waves.
4. Connection between the position of CDW dislocations with respect to the vortices of the uniform superconductor and the PDW in the Ginzburg-Landau formalism
5. CDW dislocations from PDW half vortices with disorder
6. Additional references

## Methods

**STM Measurements**

Single crystals of UTe$_2$ were used for this measurement. The growth and characterization are mentioned in detail elsewhere (20). Samples were cleaved in situ at ~90 K and in an ultrahigh-vacuum chamber. The samples were immediately transferred to the STM head after cleaving. The STM is housed in a $^3$He cryostat with a base temperature of 300 mK and a superconducting magnet that can apply magnetic fields up to 11 T. The temperature values reported were measured at the $^3$He pot; the actual sample temperature could be slightly higher. The STM using chemically etched and annealed tungsten tips.

**Extraction of the amplitude and phase maps from the topography using Fourier decomposition**

STM topographies in real space $T(r)$ can be thought of as a sum of various periodic signals pertaining to the atomic lattice and other modulations (for e.g., CDW in our case). To extract the amplitude and phase of the periodic signal, it is more convenient to shift to the Fourier space. The Fourier component of the CDW can be thought of as $\rho_{q_i}(r) = \rho_{q_i}^o(r) e^{i q_i^{meas}(r)}$, $i = 1,2,3$, where $\rho_{q_i}^o(r)$ is the amplitude of the $i$th CDW, and $q_i^{meas}(r) = (q_i r + \phi_i(r))$ where $q_i r$ is the perfectly modulating component of the CDW and $\phi_i(r)$ captures the phase meandering due to topological defects like dislocations and discommensurations.

We first perform an inverse Fourier filtering of the CDW peaks in the Fourier transform to isolate the signal associated with the CDW modulation. This signal is $\rho_{q_i}^o(r) e^{i q_i^{meas}(r)}$. To visualize the amplitude and the modulating component for the CDW in real space, we take the modulus and the cosine of the argument of this complex signal and plot those as a function of position. To plot the relative phase $\phi_i(r)$ of the CDW signal, we multiply a modulating component $e^{-i q_i(r)}$ to $\rho_{q_i}(r)$. Then the $Phase[\rho_{q_i}(r) e^{-i q_i(r)}] = Phase[\rho_{q_i}^o(r) e^{i \phi_i(r)}] = \phi_i(r)$ carries the information about the phase slips in the CDW order. So, we isolate the argument of $\rho_{q_i}(r) e^{-i q_i(r)}$ and plot it as a function of position $(r)$. This relative phase varies between -π to π. This implies that the relative phase stays largely constant in the field of view of the topography and winds from -π to π (or vice-versa) around topological defects. This process is similar to what has been described in the supplementary information of ref. (30). To find the exact pixel coordinates of the CDW peaks in the Fourier transform, we take the center of mass of a window of size of $\sigma = 0.17 \, nm^{-1}$. We take the smallest windowing size possible that captures only the CDW signal and inverse Fourier filter the signal inside this window for the above analysis. This process is similar to the standard lock-in technique followed for Fourier decomposition of CDWs and PDWs in cuprates (31,32).

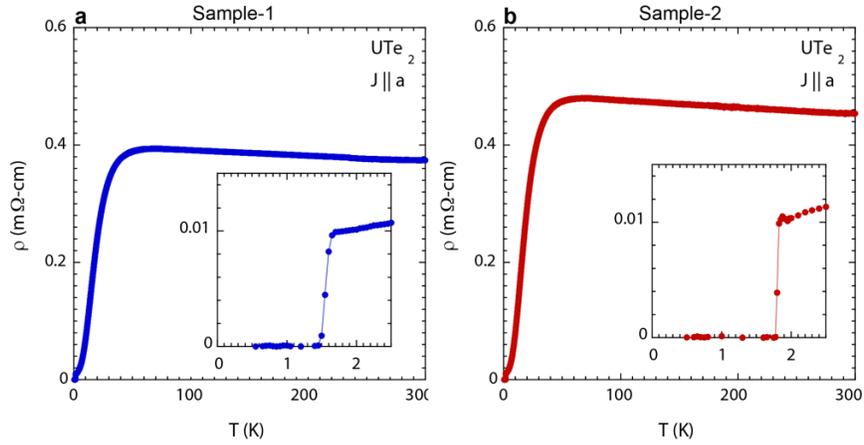

**Supplementary Figure 1: Transport characterization of the UTe$_2$ single crystals.**
**a-b,** Transport characterization of samples from the two growth batches that were used in this study. Inset shows the superconducting transition for both the samples. **a** and **b** have slightly different T$_c$ owing to their different growth conditions.

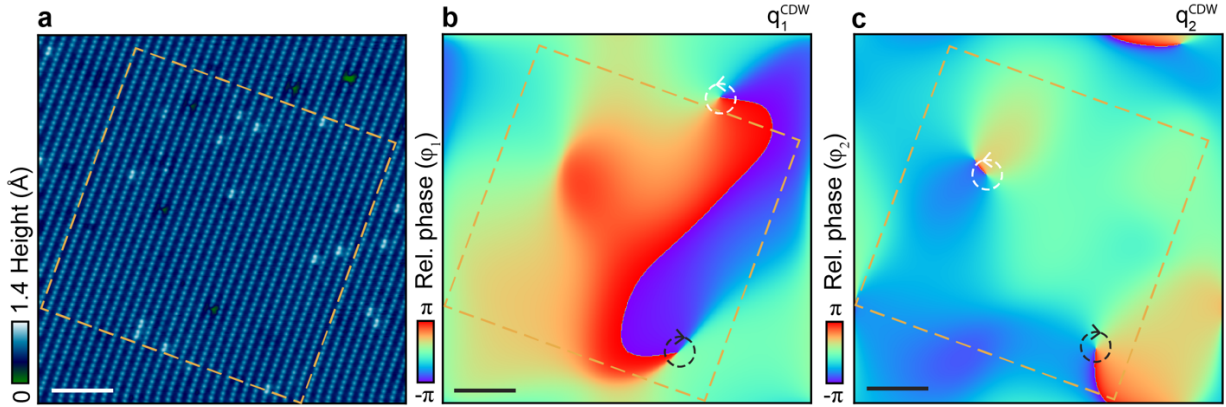

**Supplementary Figure 2: Original raw 300 Å x 300 Å topography along with relative phase maps for different CDW components**

**a**, Original high-resolution topography on the (011) surface of UTe$_2$ showing the Te1 and Te2 atoms (V = -40 mV, I = 120 pA, T = 300 mK) from which the topography shown in Figure 1a was cropped. The cropped area is marked with a dashed square. Scale bar is 50 Å. **b-c**, Relative phase maps for the two CDW components $q_1^{CDW}$ and $q_2^{CDW}$ showing the presence of topological defects. The cropped area is marked with a dashed square. Scale bars are 50 Å.

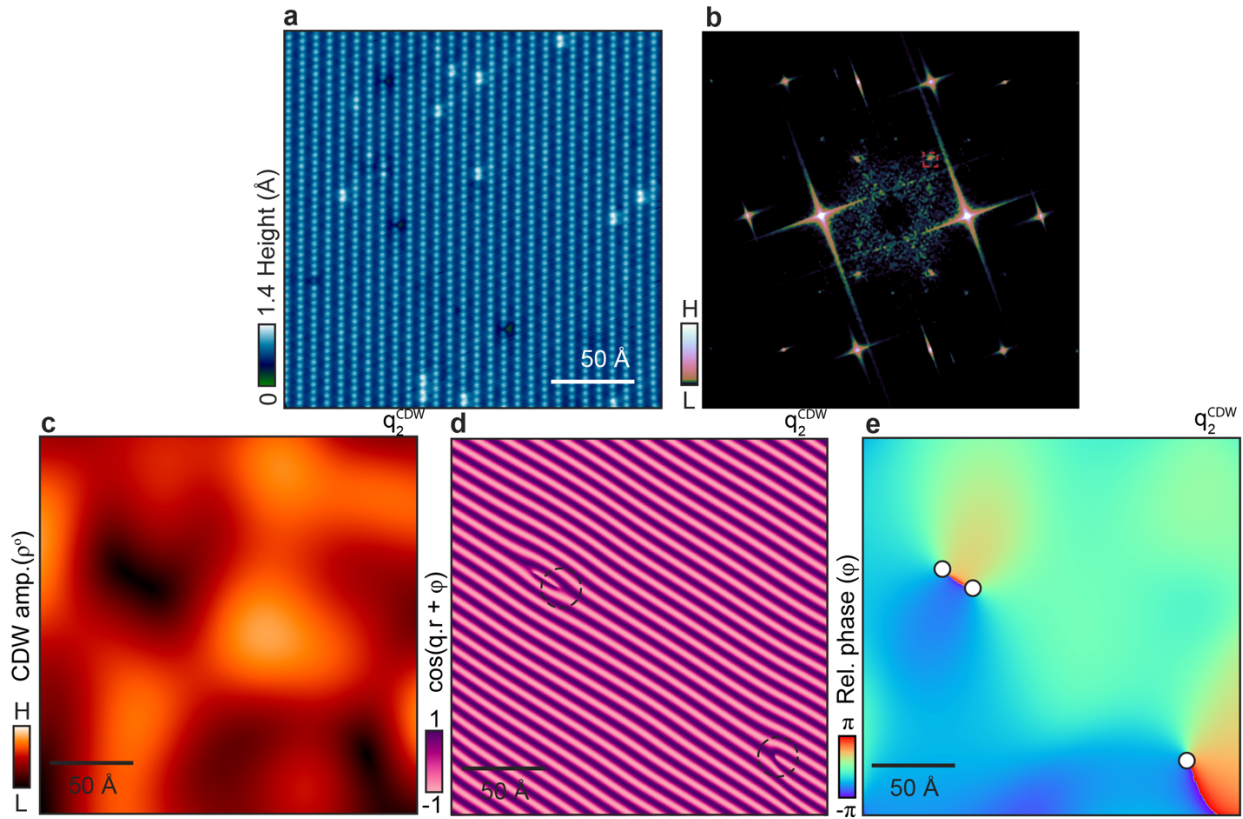

**Supplementary Figure 3: Presence of topological defects in $q_2^{CDW}$**

**a**, High resolution topography on the (011) surface of UTe$_2$ showing the Te1 and Te2 atoms (V = -40 mV, I = 120 pA, T = 300 mK). Scale bar is 50 Å. **b**, Fourier transform of the topography shown in **a**, with the $q_2^{CDW}$ CDW peak is shown in dashed red square. **c-e**, Amplitude, $cos(q_2 r + \phi_2(r))$ and relative phase maps of the order parameter associated with $q_2^{CDW}$. Scale bar is 50 Å. **e** and **f** show the position of isolated topological defects in the CDW.

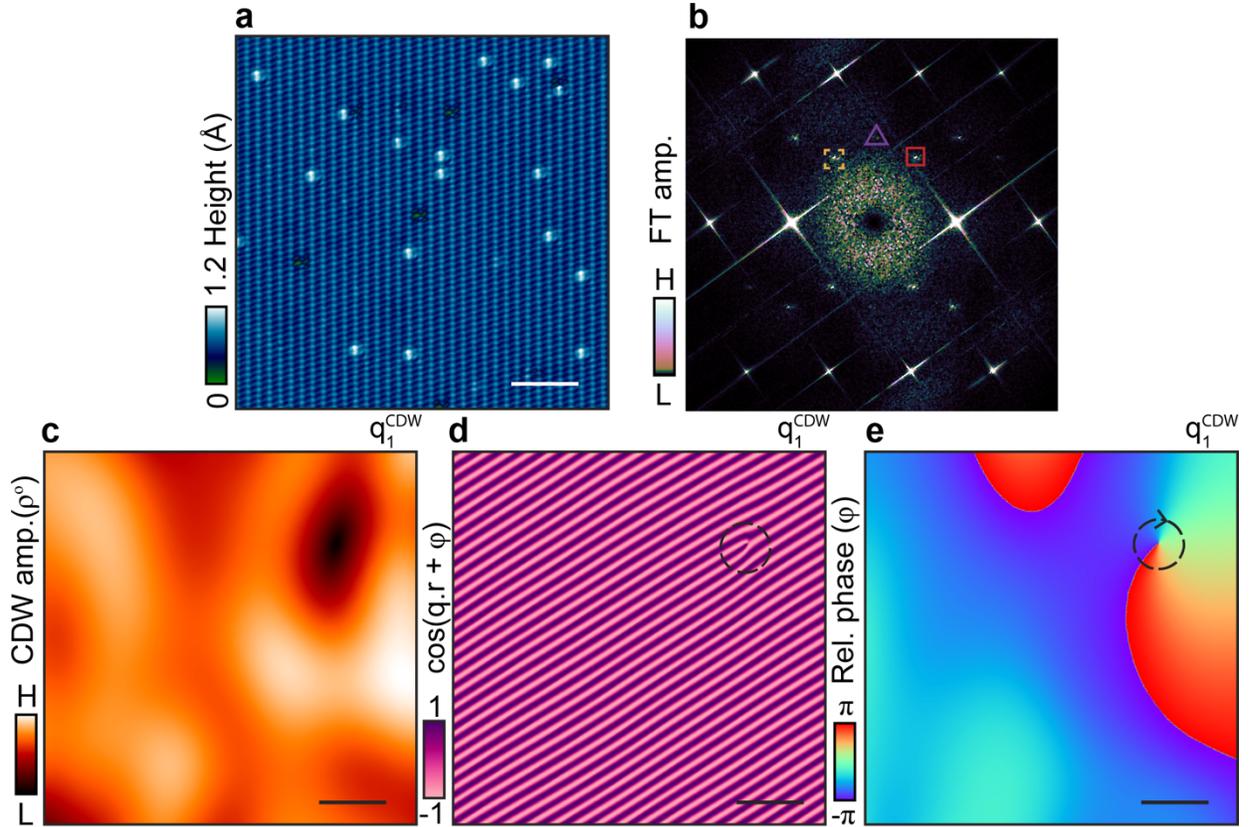

**Supplementary Figure 4: Additional dataset with a different sample and tip, that shows the presence of isolated topological defects at 0 T.**

**a**, High resolution topography on the (011) surface of UTe$_2$ showing the Te1 and Te2 atoms (V = -60 mV, I = 200 pA, T = 300 mK). Scale bar is 50 Å. **b**, Fourier transform of the topography shown in **a**, with the the CDW peaks shown by the squares and triangles. $q_1^{CDW}$ CDW peak is shown in dashed square. **c-e**, Amplitude, $cos(q_1 r + \phi_1(r))$ and relative phase maps of the order parameter associated with $q_1^{CDW}$. Scale bars are 50 Å. **d** and **e** show the position of isolated topological defects in the CDW.

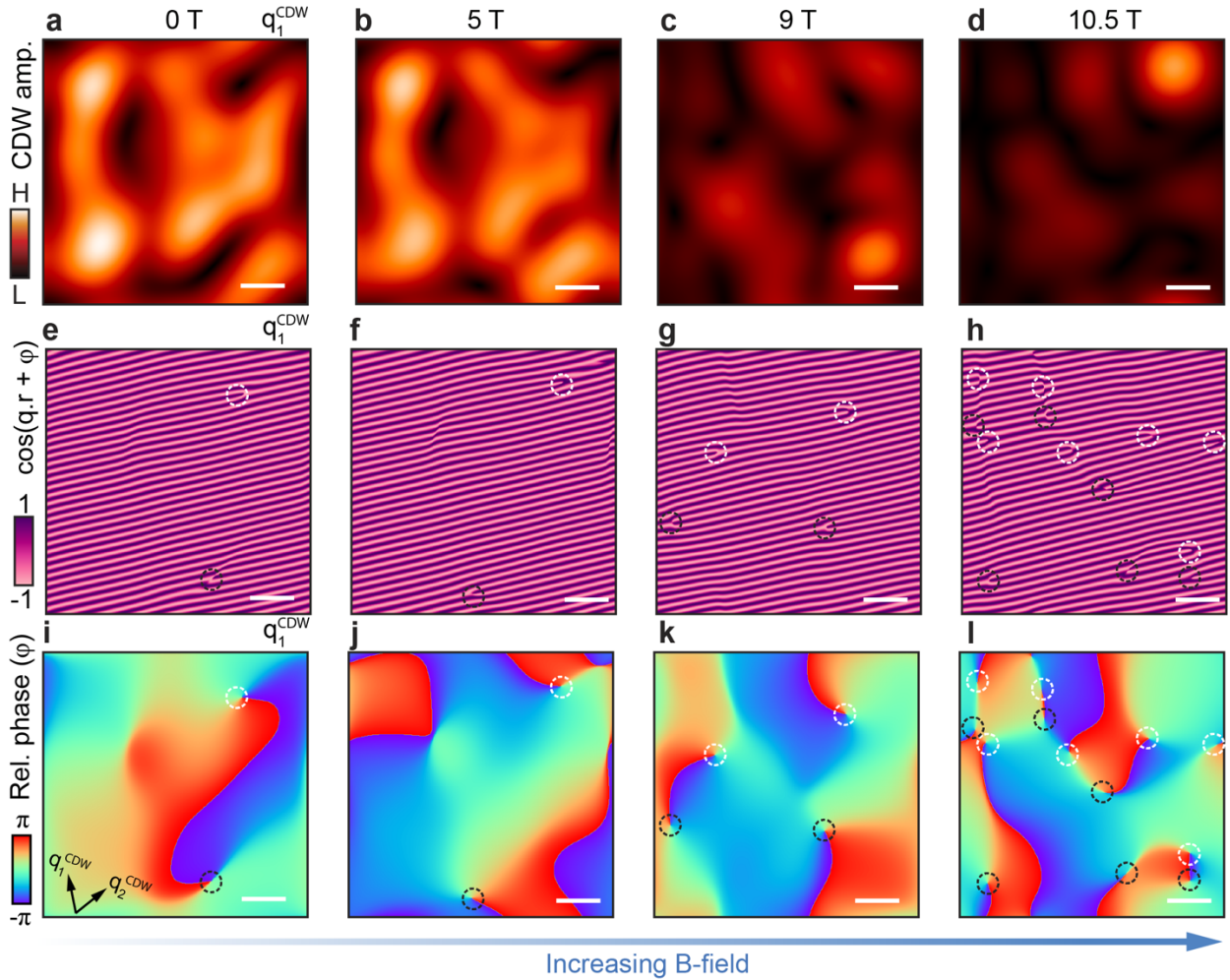

**Supplementary Figure 5: Generation of topological defects and decay of the amplitude with increasing magnetic field for $q_1^{CDW}$**

**a-d**, Suppression of amplitude shown by amplitude maps of $q_1^{CDW}$ as a function increasing magnetic field. **(e-h),(i-l)** $cos(q_1 r + \phi_1(r))$ maps and relative phase maps of $q_1^{CDW}$ (respectively) as a function of magnetic field showing an increase in vortex-antivortex pairs in the CDW phase. The vortex and anti-vortex are shown by black and white dashed circles respectively. (V = -40 mV, I = 120 pA, T = 300 mK). Scale bars are 50 Å in all maps.

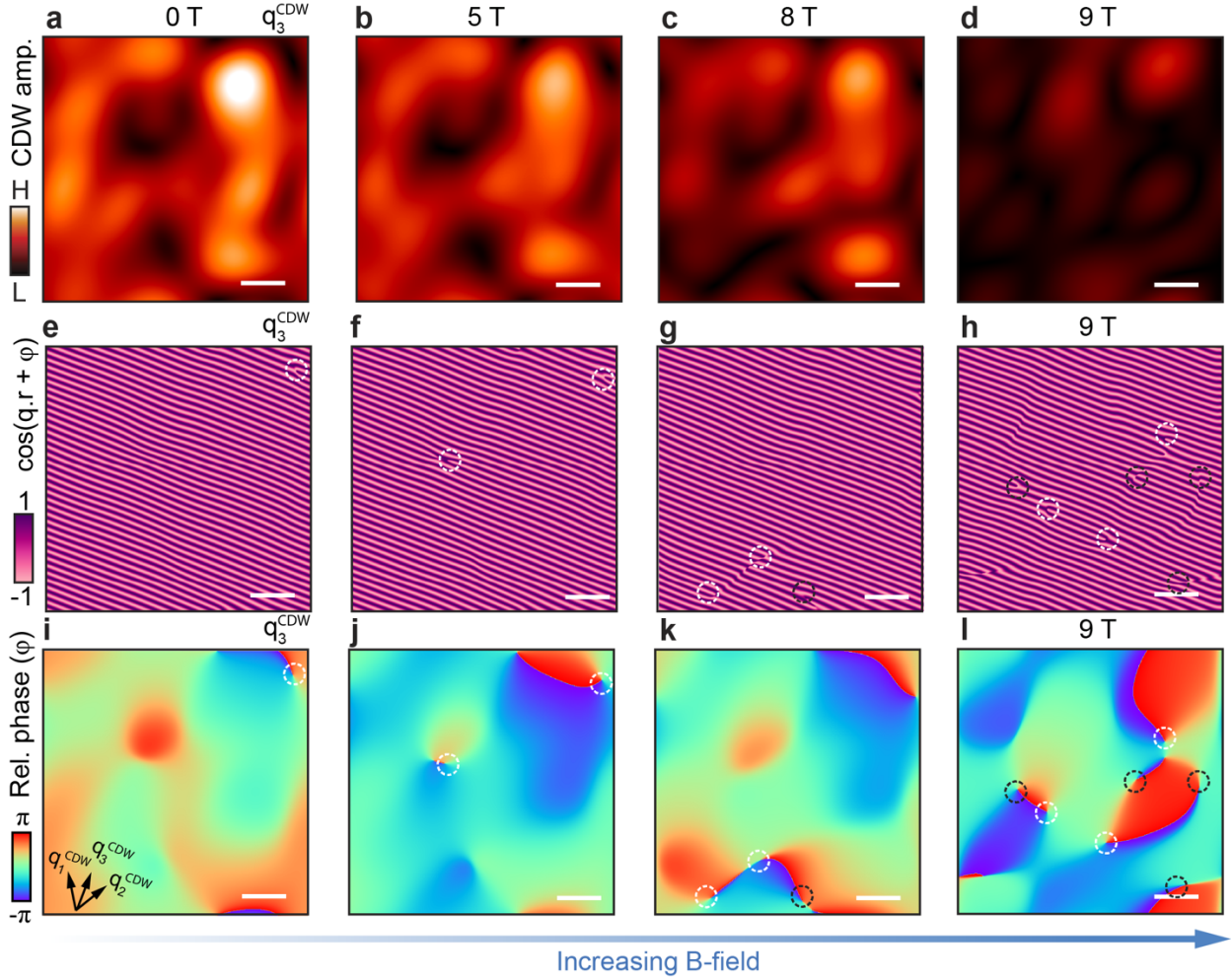

**Supplementary Figure 6: Generation of topological defects and decay of the amplitude with increasing magnetic field for $q_3^{CDW}$**

**a-d**, Suppression of amplitude shown by amplitude maps of $q_3^{CDW}$ as a function increasing magnetic field. The color scale for the amplitude of $q_3^{CDW}$ is saturated at 70% in comparison to $q_1^{CDW}$ and $q_2^{CDW}$ because of its smaller magnitude. For the same reason all the maps shown above for $q_3^{CDW}$ have been obtained from the topographies taken at positive bias. We stop at 9 T for $q_3^{CDW}$, as the signal becomes undetectable above noise level for fields above 9 T. **(e-h),(i-l)** $cos(q_3 r + \phi_3(r))$ maps and relative phase maps of $q_3^{CDW}$ (respectively) as a function of magnetic field showing an increase in vortex-antivortex pairs in the CDW phase. The vortex and anti-vortex are shown by black and white dashed circles respectively. (V = 40 mV, I = 120 pA, T = 300 mK). Scale bars are 50 Å in all maps.

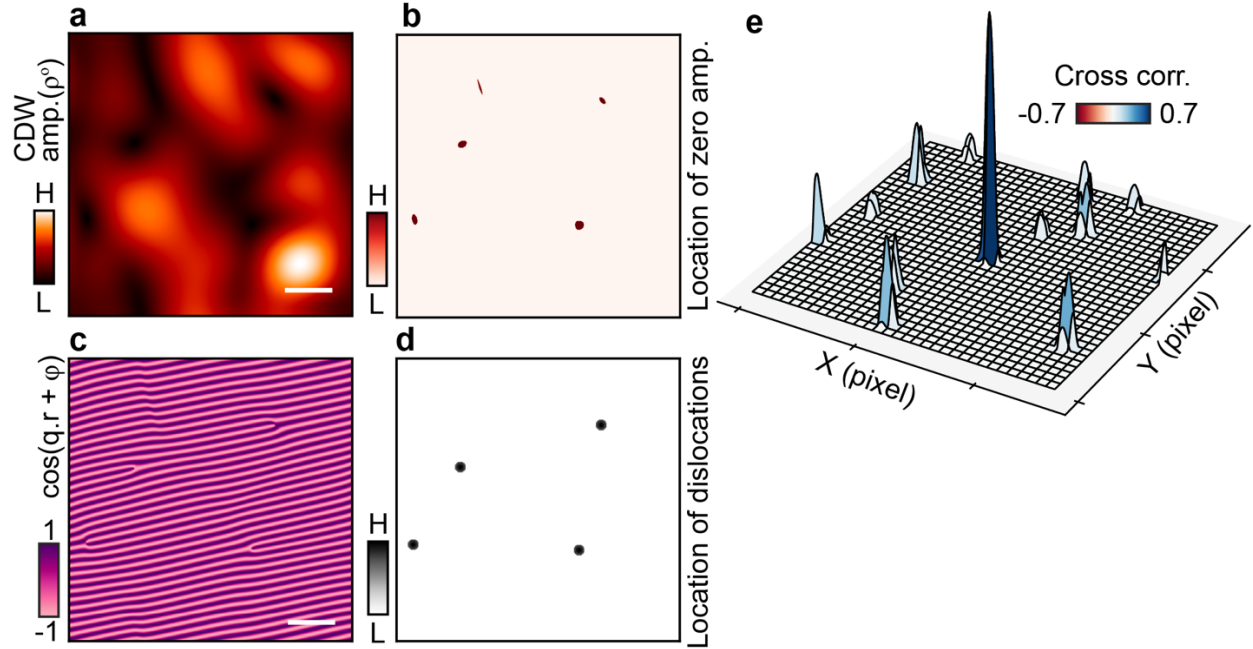

**Supplementary Figure 7: Cross-correlation between regions of zero-amplitude and location of the topological defects for $q_1^{CDW}$**

**a**, Amplitude map of $q_1^{CDW}$ obtained at B = 9 T, T = 300 mK. **b**, Zero-amplitude mask generated by a 5% thresholding from the amplitude map which shows regions of zero amplitude. All scale bars are 50 Å. **c**, $cos(q_2 r + \phi_2(r))$ map of $q_1^{CDW}$ at B = 9 T. **d**, Binary mask of the position of dislocations. Each dislocation is represented by a circular gaussian function. **e**, 2D cross-correlation of the normalized masks shown in **b** and **d** with a value 0.6 at the center indicative of a positive cross-correlation.

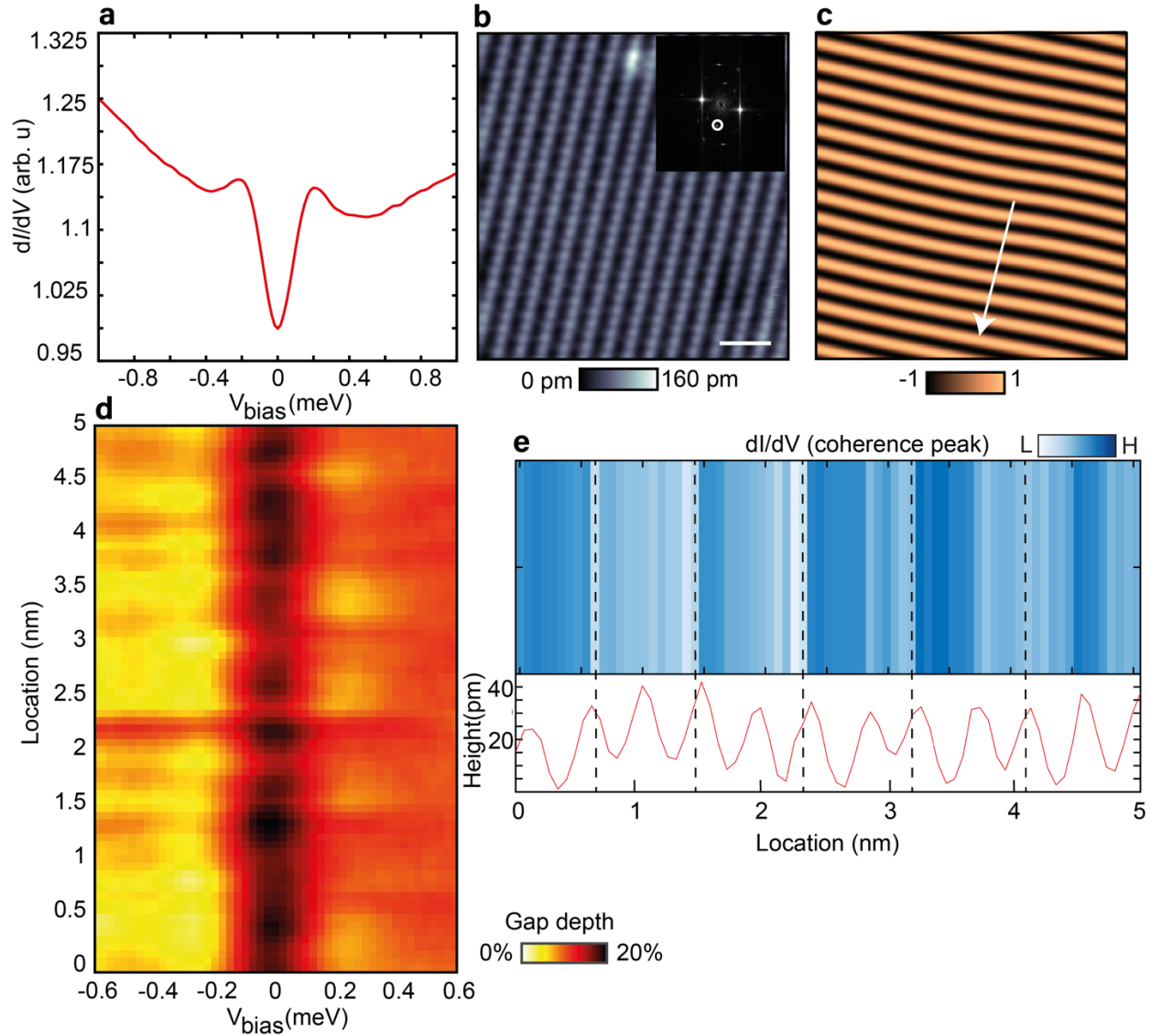

**Supplementary Figure 8: Spectroscopic evidence of superconductivity and location dependence of superconducting gap**

**a**, Average d$I$/d$V$ spectrum obtained on the (011) surface at T = 300 mK. (V = 2 mV, I = 50 pA, $V_{mod}$ = 50 μeV). **b**, Topography obtained at 300 mK on which the dI/dV spectra shown in **a** and **c** have been obtained. (V = 15 mV, I = 50 pA) Scale bar is 20 Å. Inset shows the FT of the topography showing the CDW peaks. **c**, Inverse Fourier filtering of the CDW peak showing the CDW modulations in real space. White arrow shows the location where the spectra in **d** were obtained. **d**, Linecut of d$I$/d$V$ spectra obtained along the white arrow indicated in **d** showing the dependence of the superconducting gap size and depth as a function of location. (V = 2 mV, I = 50 pA, $V_{mod}$ = 50 μeV). **e**, Plot of coherence peak oscillation as a function of position. The panel below shows the height profile showing atomic corrugation. The coherence peak oscillates with periodicity close to twice the atomic lattice corresponding to the incommensurate CDW and PDW.

**Ginzburg-Landau theory for coexisting uniform triplet superconductivity and triplet pair density waves.**

The Ginzburg-Landau free energy for a system with uniform triplet superconducting and triplet pair density wave (PDW) orders can be expanded in terms of the order parameters, $\mathbf{\Delta}_0$ and $\mathbf{\Delta}_{\pm q_i}$ respectively, where $q_i$ are the wavevectors of the PDW orders. For our purposes, we consider the following "minimal" free energy needed to capture the ordered phase,

$$F = \kappa_0 |D_\mu \mathbf{\Delta}_0|^2 + \sum_i \kappa_i \left[ |D_\mu \mathbf{\Delta}_{+q_i}|^2 + |D_\mu \mathbf{\Delta}_{-q_i}|^2 \right]$$

$$+ r_0 |\mathbf{\Delta}_0|^2 + \sum_i r_{q_i} \left[ |\mathbf{\Delta}_{+q_i}|^2 + |\mathbf{\Delta}_{-q_i}|^2 \right] + u_0 |\mathbf{\Delta}_0|^4 + \sum_i u_{q_i} \left[ |\mathbf{\Delta}_{+q_i}|^4 + |\mathbf{\Delta}_{-q_i}|^4 \right]$$

$$+ \sum_i \alpha_i \; \mathbf{\Delta}_0 \cdot \mathbf{\Delta}_0 \; \mathbf{\Delta}^*_{+q_i} \cdot \mathbf{\Delta}^*_{-q_i} + \sum_i \alpha'_i \; \mathbf{\Delta}_0 \cdot \mathbf{\Delta}^*_{+q_i} \; \mathbf{\Delta}_0 \cdot \mathbf{\Delta}^*_{-q_i} + \sum_{ij} \beta_{ij} \; \mathbf{\Delta}_{+q_i} \cdot \mathbf{\Delta}_{-q_i} \; \mathbf{\Delta}^*_{+q_j} \cdot \mathbf{\Delta}^*_{-q_j}$$

$$+ \sum_{ij} \beta'_{ij} \; \mathbf{\Delta}_{+q_i} \cdot \mathbf{\Delta}^*_{+q_j} \; \mathbf{\Delta}_{-q_i} \cdot \mathbf{\Delta}^*_{-q_j} + \sum_{ij} \beta''_{ij} \; \mathbf{\Delta}_{+q_i} \cdot \mathbf{\Delta}^*_{-q_j} \; \mathbf{\Delta}_{-q_i} \cdot \mathbf{\Delta}^*_{+q_j} + h.c.. + \cdots$$

where we are enforcing inversion symmetry, which sends $\mathbf{\Delta}_{+q_i} \leftrightarrow \mathbf{\Delta}_{-q_i}$. The '…' stands in for other symmetry allowed term that we have omitted for brevity. This notably includes the bi-quadratic terms proportional to $|\mathbf{\Delta}_0|^2 |\mathbf{\Delta}_{\pm q_i}|^2$ and $|\mathbf{\Delta}_{\pm q_i}|^2 |\mathbf{\Delta}_{\pm q_j}|^2$. These terms are important to understanding the full structure of the ordered phases, but they will not be directly relevant to our forthcoming discussion. The phase with coexisting triplet superconductivity and triplet PDWs occurs when $r_0, r_{q_i} < 0$ and $u_0, u_{q_i} > 0$.

The phase with coexisting uniform superconducting and PDW orders has a daughter charge density wave (CDW) order with the same wavevector as the PDW order. Within the Ginzburg-Landau theory, the daughter CDW order arises from the following terms in the free energy,

$$F_{CDW} = \sum_i c_{q_i} |\rho_{+q_i}|^2 + \sum_i \lambda_i \, \rho^*_{+q_i} [\, \mathbf{\Delta}_0 \cdot \mathbf{\Delta}^*_{-q_i} + \mathbf{\Delta}^*_0 \cdot \mathbf{\Delta}_{+q_i}] + h.c.,$$

where $\rho_{+q_i}$ is the order parameter for the CDW with wavevector $q_i$, and $\rho^*_{+q_i} = \rho_{-q_i}$ since the CDWs are real order parameters. For brevity, we are again omitting terms from the free energy that are unimportant for our current discussions. When $c_{q_i} > 0$, the CDW is a daughter order to the uniform superconducting and PDW orders. From the equations of motion for $\rho^*_{q_i}$ we have that

$$\rho_{+q_i} = -\frac{\lambda_i}{c_{q_i}} \left[ \mathbf{\Delta}_0 \cdot \mathbf{\Delta}^*_{-q_i} + \mathbf{\Delta}^*_0 \cdot \mathbf{\Delta}_{+q_i} \right].$$

The CDW order parameter $\rho_{+q_i}$ therefore only has an expectation value when $\mathbf{\Delta}_0$ and $\mathbf{\Delta}_{\pm q_i}$ have expectation values.

**Connection between CDW dislocations and vortices of the uniform superconductor and the PDW**

In the main text we showed that it is possible for half-vortices of the PDW order parameter to produce dislocations of the daughter CDW. Here, we will show that vortices of the uniform superconductivity can also produce dislocations, provided that there is additional inversion symmetry breaking.

Let us consider a vortex of the uniform superconductor, were $\Delta_0 \propto e^{i\theta_v}$ ; $\theta_v$ winds by $2\pi$ around the vortex core; and $\Delta_{\pm q_i}$, are constant near the vortex. Based on the equations of motion for $\rho_{+q_i}$, the vortex of $\Delta_0$ will produce a dislocation when $|\Delta_0^* \cdot \Delta_{+q_i}| < |\Delta_0 \cdot \Delta_{-q_i}^*|$ away from the vortex core, and an anti-dislocation when $|\Delta_0^* \cdot \Delta_{+q_i}| > |\Delta_0 \cdot \Delta_{-q_i}^*|$. However, it is expected that $|\Delta_0^* \cdot \Delta_{+q_i}|=|\Delta_0 \cdot \Delta_{-q_i}^*|$ away from the vortex core due to inversion symmetry. In this case it is impossible to satisfy the condition to form either a dislocation or anti-dislocation. It is therefore only possible, for a vortex of the uniform superconducting order parameter to produce a dislocation or anti-dislocation if there is additional inversion symmetry breaking. This is expected on symmetry grounds. This consideration does not apply when considering a single half-vortex of the PDW, as the existence of a half-vortex already breaks inversion symmetry, as a half-vortex is a phase winding of $\Delta_{+q_i}$ but not $\Delta_{-q_i}$ or vice versa.

**CDW dislocations from PDW half vortices with disorder**

To explain the existence of the observed CDW dislocations in zero magnetic field, we will consider adding quenched random disorder () to the Ginzburg-Landau theory of the daughter CDW state. In systems with random disorder that break translational symmetry, we can include the following term in the free energy for the CDW state,

$$F_{dis} = \sum_i v_i \rho_{+q_i} + h.c.,$$

where $v_i$ is complex random potential that originate from the disorder in the system. Disorder can also cause the parameters in the free energy to vary in space, but we will ignore such effects for this analysis. Including $F_{dis}$, the equations of motion for the CDW order parameter are

$$\rho_{+q_i} = -\frac{\lambda_i}{c_{q_i}} \left[ \Delta_0 \cdot \Delta_{-q_i}^* + \Delta_0^* \cdot \Delta_{+q_i} \right] - v_i^*.$$

We can now discuss how topological defects of $v_i$ where the phase winds by $\pm 2\pi$ can produce anti-dislocations or dislocations of the CDW respectively.

Let us first consider the $2\pi$ phase windings of $v_i$. Specifically, let us consider the case where $v_i \propto e^{i\theta_v}$ ; $\theta_v$ winds by $2\pi$ around the core of the defect; and $\Delta_0$, and $\Delta_{\pm q_i}$ are constant. From the equations of motion, this will lead to an anti-dislocation if $|v_i^*| > \left|\frac{\lambda_i}{c_{q_i}} \left[ \Delta_0 \cdot \Delta_{-q_i}^* + \Delta_0^* \cdot \Delta_{+q_i} \right]\right|$ outside of the core of the defect of $v_i$. Similarly, a $-2\pi$ phase windings of $v_i$ will produce a dislocation if $|v_i^*| > \left|\frac{\lambda_i}{c_{q_i}} \left[ \Delta_0 \cdot \Delta_{-q_i}^* + \Delta_0^* \cdot \Delta_{+q_i} \right]\right|$ outside of the core of the defect of $v_i$. If $|v_i^*| < \left|\frac{\lambda_i}{c_{q_i}}\left[\boldsymbol{\Delta}_0 \cdot \boldsymbol{\Delta}_{-q_i}^* + \boldsymbol{\Delta}_0^* \cdot \boldsymbol{\Delta}_{+q_i}\right]\right|$ then then the phase windings of $v_i$ will not produce any dislocations or anti-dislocations. As expected, we find that quenched random disorder can create dislocations and anti-dislocations of the CDW, even the absence of any magnetic field induced half-vortices of the PDW.

It is also useful to revisit how half-vortex of $\boldsymbol{\Delta}_{+q_i}$ can produce dislocations of the CDW if we include disorder. Let us consider a half-vortex where $\boldsymbol{\Delta}_{+q_i} \propto e^{i\theta_v}$; $\theta_v$ winds by $2\pi$ around the core of the half-vortex; and $\boldsymbol{\Delta}_0$, $\boldsymbol{\Delta}_{-q_i}$, and $v_i$ are constant. If $|\boldsymbol{\Delta}_0^* \cdot \boldsymbol{\Delta}_{+q_i}| > |\boldsymbol{\Delta}_0 \cdot \boldsymbol{\Delta}_{-q_i}^* + \frac{c_{q_i}}{\lambda_i}v_i^*|$ away from the half-vortex core, the half-vortex of $\boldsymbol{\Delta}_{+q_i}$ will lead to a dislocation of $\rho_{+q_i}$. Additionally, if $|\boldsymbol{\Delta}_0^* \cdot \boldsymbol{\Delta}_{+q_i}| < |\boldsymbol{\Delta}_0 \cdot \boldsymbol{\Delta}_{-q_i}^* + \frac{c_{q_i}}{\lambda_i}v_i^*|$ there will not be a dislocation. Similar reasoning indicates that a half-vortex of $\boldsymbol{\Delta}_{-q_i}$ will lead to an anti-dislocation of $\rho_{+q_i}$, if $|\boldsymbol{\Delta}_0 \cdot \boldsymbol{\Delta}_{-q_i}^*| > |\boldsymbol{\Delta}_0^* \cdot \boldsymbol{\Delta}_{+q_i} + \frac{c_{q_i}}{\lambda_i}v_i^*|$ away from the half-vortex core. Interestingly, from this analysis we find that the disorder, $v_i$ can either stabilize or destabilize the dislocations induced by the half-vortices, even when $v_i^*$ does not have any topological defects. This is especially important if we consider the case where $|\boldsymbol{\Delta}_0^* \cdot \boldsymbol{\Delta}_{+q_i}| = |\boldsymbol{\Delta}_0 \cdot \boldsymbol{\Delta}_{-q_i}^*|$ (as would be expected from inversion symmetry) away from the half-vortex cores.

As a final point, we revisit the possibility of having dislocations induced by vortices of the uniform superconductivity. Similar to before, we find that vortices of the uniform superconductivity can only produce dislocations or anti-dislocations if there is additional inversion symmetry breaking. Let us consider a vortex of the uniform superconductor, were $\boldsymbol{\Delta}_0 \propto e^{i\theta_v}$; $\theta_v$ winds by $2\pi$ around the vortex core; and $\boldsymbol{\Delta}_{\pm q_i}$, and $v_i$ are constant. The vortex of $\boldsymbol{\Delta}_0$ will produce a dislocation when $|\boldsymbol{\Delta}_0^* \cdot \boldsymbol{\Delta}_{+q_i} + \frac{c_{q_i}}{\lambda_i}v_i^*| < |\boldsymbol{\Delta}_0 \cdot \boldsymbol{\Delta}_{-q_i}^*|$ away from the vortex core, and an anti-dislocation when $|\boldsymbol{\Delta}_0^* \cdot \boldsymbol{\Delta}_{+q_i}| > |\boldsymbol{\Delta}_0 \cdot \boldsymbol{\Delta}_{-q_i}^* + \frac{c_{q_i}}{\lambda_i}v_i^*|$ away from the vortex core. However, it is expected that $|\boldsymbol{\Delta}_0^* \cdot \boldsymbol{\Delta}_{+q_i}| = |\boldsymbol{\Delta}_0 \cdot \boldsymbol{\Delta}_{-q_i}^*|$ away from the vortex core due to inversion symmetry. It is then impossible to satisfy the condition to form either a dislocation or anti-dislocation.

**Additional references**